# THE IMPORTANCE OF THE ALGORITHMIC INFORMATION THEORY TO CONSTRUCT A POSSIBLE EXAMPLE WHERE NP ≠ P

Rubens Viana Ramos

rubens@deti.ufc.br

*Department of Teleinformatic Engineering – Federal University of Ceara - DETI/UFC*

*C.P. 6007 – Campus do Pici - 60755-640 Fortaleza-Ce Brazil*

In this short communication it is shown a simple problem using quantum circuits for which the algorithmic information theory guarantee that the minimal length of the algorithm able to solve it grows exponentially with the number of qubits.

In Fig. 1 it is presented a *n*-qubit quantum circuit composed by $n(n+1)$ single-qubit gates and $n(n-1)$ CNOTs [1]. A lot of different $2^n \times 2^n$ unitary matrices can be produced using the quantum circuit of Fig. 1 enabling or not the CNOTs and setting values for the single-qubit gates (to be more precise, a set of three angles values that I take as a unique value for the quantum gate).

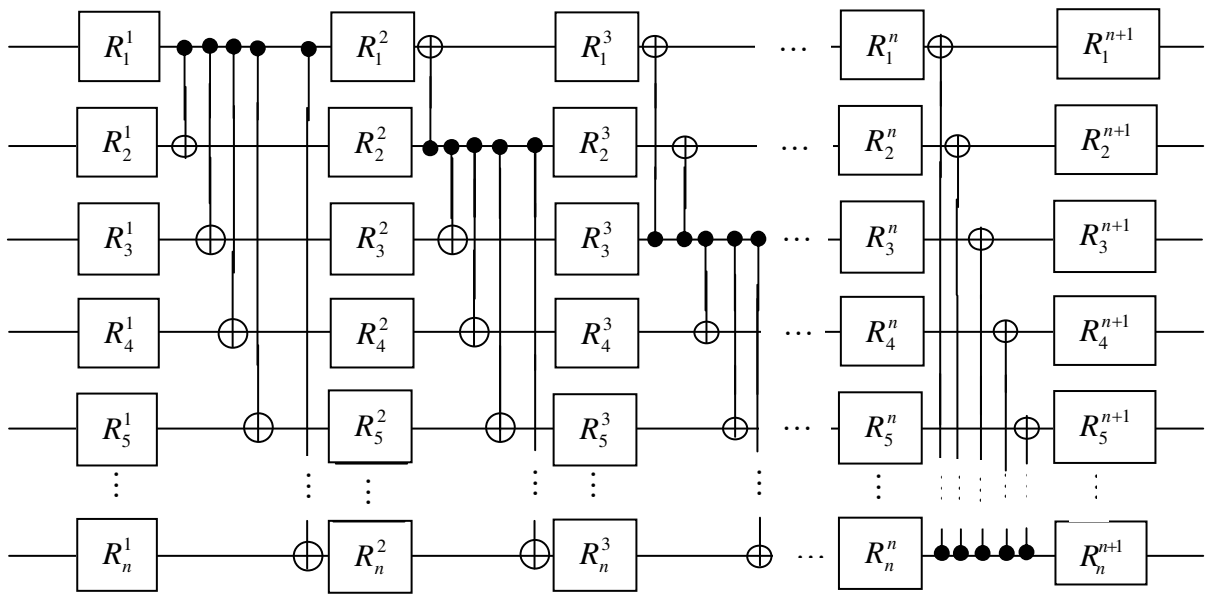

Figure 1: *n*-qubit quantum circuit. *R* are single-qubit gates.

Let us assume that each single-qubit gate value shall be chosen from a set of only *k* different values or, in other words, there exist only *k* different types of single-qubit gate allowed. These values are properly chosen in order to avoid different configurations providing the same total $2^n \times 2^n$ unitary matrix. This is not very hard to achieve since there are infinite values for the single-qubit gates to be tested. Under these conditions, using the quantum circuit of Fig.1 for *n* qubits, it is possible to construct $k^{[n(n+1)]} 2^{[n(n-1)]}$ different $2^n \times 2^n$ unitary matrices.

Now, suppose a communication between two distant parts, Alice and Bob. Alice uses the quantum circuit of Fig. 1 to produce $k^{[n(n+1)]} 2^{[n(n-1)]}$ different unitary matrices, she sends them to Bob and ask him to create a computer program that provides as output the matrices sent by her. Bob knows that Alice is using the quantum circuit of Fig 1 but he has no idea about the single-qubit gates she has used, hence, the set of matrices sent by Alice seems to Bob a set of unitary matrices randomly chosen (there is not a formation law to generated all the matrices sent by Alice). In this case, the algorithmic information theory [2] states that the best that Bob can do is to create a program that takes the matrices sent by Alice as defined variables and then print them. Hence, the computer program with minimal length (number of bits) able to provide as output all the matrices sent by Alice must contain all those matrices. However, since the number of matrices grows exponentially with the number of qubit *n*, the minimal length of the program also grows exponentially with *n*. Since this program has already the minimal length, it is not possible to construct, by any method, any other program whose length grows only in a polynomial way.